\documentclass[aps,prd,nofootinbib,showkeys,showpacs,floatfix]{revtex4}
\usepackage{amsmath}
\usepackage{amstext}
\usepackage{amssymb}
\usepackage{graphicx}
\usepackage{hyperref}
\usepackage{rotating}
\usepackage{gensymb}
\usepackage{color} 
\usepackage{multirow} 
\usepackage[T1]{fontenc} 
\usepackage{slashed}

\def\gsim{\raise0.3ex\hbox{$\;>$\kern-0.75em\raise-1.1ex\hbox{$\sim\;$}}}
\def\lsim{\raise0.3ex\hbox{$\;<$\kern-0.75em\raise-1.1ex\hbox{$\sim\;$}}}

\newcommand{\eVq}  {\text{eV}^2}

\newcommand{\AddrAHEP}{%
  AHEP Group, Institut de F\'{i}sica Corpuscular --
  C.S.I.C./Universitat de Val\`{e}ncia, Parc Cientific de Paterna.\\
 C/ Catedratico Jos\'e Beltr\'an, 2 E-46980 Paterna (Val\`{e}ncia) - SPAIN}

\begin{document}


\title{Neutrino oscillations refitted}

\author{D.~V.~Forero} \email{dvanegas@ific.uv.es} \affiliation{\AddrAHEP}
\author{M. T{\'o}rtola}\email{mariam@ific.uv.es}    \affiliation{\AddrAHEP}
\author{J.~W.~F.~Valle} \email{valle@ific.uv.es} \affiliation{\AddrAHEP}

\vskip 2cm

\begin{abstract}

  Here we update our previous global fit of neutrino oscillations by
  including the recent results which have appeared since the
  Neutrino-2012 conference. These include the measurements of reactor
  anti-neutrino disappearance reported by Daya Bay and RENO, together
  with latest T2K and MINOS data including both disappearance and
  appearance channels.
  We also include the revised results from the third solar phase of 
  Super-Kamiokande, SK-III, as well as new solar results from the
  fourth phase of Super-Kamiokande, SK-IV.
  We find that the preferred global determination of the atmospheric angle 
  $\theta_{23}$ is consistent
  with maximal mixing. We also determine the impact of the new data
  upon all the other neutrino oscillation parameters with emphasis on
  the increasing sensitivity to the CP phase, thanks to the interplay
  between accelerator and reactor data.
In the appendix we present the updated results obtained after the
inclusion of new reactor data presented at the Neutrino-2014
conference. We discuss their impact on the global neutrino analysis.
\end{abstract}
\maketitle

\section{Introduction}

The precise measurement of a non-zero value of the third mixing angle
$\theta_{13}$ in the lepton mixing
matrix~\cite{schechter:1980gr,Beringer:1900zz,Rodejohann:2011vc}
reported by the reactor experiments Daya Bay~\cite{An:2012eh} and
RENO~\cite{Ahn:2012nd} has played an important role in electroweak
model-building as well as in the design of future upcoming experiments
over the last two years.
Compared to previous reactor anti-neutrino experiments
CHOOZ~\cite{Apollonio:2002gd} and Palo Verde~\cite{Boehm:2001ik},
these new measurements have larger statistics, thanks to the increased
reactor power and anti-neutrino detector size involved.  More
importantly, they have detectors located at different distances from
the reactor cores, in order to reduce the effect of the systematic
uncertainties, such as the ones associated to the predicted
theoretical reactor fluxes.
As a result these experiments have been able for the first time to
observe the disappearance of reactor anti-neutrinos over short
distances, of the order of 1~km, providing the first robust
measurement of the mixing angle $\theta_{13}$.
Moreover, there have also been indications for a non-zero
$\theta_{13}$ mixing angle coming from the observation of electron
neutrino appearance on a muon neutrino beam at the accelerator
oscillation experiments T2K~\cite{Abe:2011sj} and
MINOS~\cite{Adamson:2011qu}.

Here we update the global fit of neutrino oscillations given in
Ref.~\cite{Tortola:2012te} by including the recent measurements of
reactor anti-neutrino disappearance reported by the Daya Bay and RENO
experiments~\cite{An:2013zwz,dyb-TAUP:13,reno-TAUP:13}, as well as
accelerator appearance and disappearance results from MINOS and
T2K~\cite{Adamson:2013whj,Adamson:2013ue,Abe:2014ugx,Abe:2013hdq}.
Concerning the solar neutrino data, our analysis includes the recently
revised results of the third solar phase of Super-Kamiokande,
SK-III~\cite{Abe:2010hy} as well as the latest results from its fourth
solar phase, SK-IV, with lower energy threshold and improved
systematic uncertainties~\cite{Renshaw:2014awa}.
We investigate the impact of the new data upon all the neutrino
oscillation parameters, discussing in more detail the status of the
octant-determination of the atmospheric mixing angle, as well as the
improved sensitivity to the CP phase $\delta$ that follows from the
complementarity of accelerator and reactor neutrino data.

\section{Updated global fit  May 2014}

\subsection*{Updated solar neutrino analysis}

As in our previous global fit to neutrino
oscillations~\cite{Tortola:2012te}, here we consider the most recent
results from the solar experiments Homestake~\cite{Cleveland:1998nv},
Gallex/GNO~\cite{Kaether:2010ag}, SAGE~\cite{Abdurashitov:2009tn},
Borexino~\cite{Bellini:2011rx},
SNO~\cite{aharmim:2008kc,Aharmim:2009gd} and the first three solar
phases of
Super-Kamiokande~\cite{hosaka:2005um,Cravens:2008aa,Abe:2010hy}.
Here we have included the revised results from the third solar phase
of Super-Kamiokande, published in December 2012 in the arXiv version
of Ref.~\cite{Abe:2010hy}. This revision corrects the estimated
systematic error on the total flux observed in Super-Kamiokande as
well as the total $^8$B flux calculation. We find that the changes
are very small and their impact on the determination of solar
oscillation parameters is hardly noticeable.
We also include the results from the fourth solar phase of
Super-Kamiokande, SK-IV~\cite{Renshaw:2014awa}. This data release
corresponds to 1306.3 live-days and is presented in the form of 23 day
and night energy bins. Thanks to several improvements in the hardware
and software of Super-Kamiokande, an improved systematic uncertainty
as well as a very low energy threshold of 3.5 MeV have been
achieved. As we will discuss later, these new data consolidate the
previous Super-Kamiokande solar data releases, with a minor impact in
the global fit to neutrino oscillations.
More detailed information on our simulation and analysis of solar
neutrino data can be found in
Refs.~\cite{Tortola:2012te,Schwetz:2011qt,Schwetz:2011zk}.

\subsection*{New reactor data}

For the statistical analysis of reactor data we follow the same
strategy as in our previous paper~\cite{Tortola:2012te}. We define a
$\chi^2$ that compares the observed and measured event rates at each
anti-neutrino detector. Several pull parameters are introduced in
order to account for the different systematical errors associated to
the reactor, detector and background uncertainties. An absolute
normalization factor is left free in the fit, to be determined from
the experimental data.
This technique is also used in the official analyses performed by the
Daya Bay and RENO Collaborations~\cite{An:2012bu,Ahn:2012nd}.
For the analysis of reactor data we take into account the total rate
analysis of the latest Double Chooz data in Ref.\cite{Abe:2012tg},
already discussed in our previous fit, as well as the new reactor data
released by Daya Bay and RENO and described below.

\subsubsection*{Daya Bay}

Daya Bay is a reactor experiment with six anti-neutrino detectors,
arranged in three experimental halls (EHs). Two detectors, located in
EH1, one in EH2 and three in EH3.  EH1 and EH2 are considered as near
detectors, while EH3 is the far detector. Electron anti-neutrinos are
generated in six reactor cores, distributed in pairs, with equal
thermal power (P$_{\rm th}^{\rm r}$ = 2.9 GW$_{\rm th}$) and detected in the EHs. The
effective baselines are 512 m and 561 m for the near halls and
1579 m for the far~\cite{An:2013zwz}. With baseline $\sim$ km
Daya Bay is sensitive to the first dip in the $\bar\nu_e$
disappearance probability.  Using this near--far technique Daya Bay has
minimized the systematic errors thus providing the most precise
determination of the reactor mixing angle $\theta_{13}$.
In Refs.~\cite{An:2013zwz,dyb-TAUP:13} Daya Bay reported $217$ days of
data collected. Such high statistics sample leads to a substantial
improvement in the statistical errors compared to the previous
analysis in Ref.~\cite{An:2012bu}. Using the corresponding event rates
at the six anti-neutrino detectors we obtain an improved measurement
of the reactor mixing angle.

\subsubsection*{RENO} 

The Reactor Experiment for Neutrino Oscillations (RENO) is a short
baseline reactor neutrino oscillation experiment located in South
Korea.
RENO consists of six reactor cores with maximum powers ranging from
2.66 GW$_\text{th}$ to 2.8 GW$_\text{th}$ and two identical
anti-neutrino detectors located at 294 and 1383 m from the center of
the reactor array.
With both near and far detectors, RENO provided an important
confirmation of the first Daya Bay measurement of
$\theta_{13}$~\cite{Ahn:2012nd}.
We use their updated results presented at the TAUP 2013
conference~\cite{reno-TAUP:13}, consisting of 403 days of data-taking,
with improved systematic uncertainties, background estimates and
energy calibration.

\subsection*{New long--baseline neutrino data}

Over the last two years new data on $\nu_\mu$ disappearance and
$\nu_e$ appearance have been released by the long-baseline (LBL) accelerator
experiments MINOS and T2K. Below we summarize the most recent data
from both experiments included in our global fit.
As in our previous analysis, we use the GLOBES software
package~\cite{Huber:2007ji} for the simulation and statistical
analysis of accelerator neutrino oscillation data from MINOS and T2K.
The expected event numbers for a given channel in a particular
detector are determined using the full three-neutrino survival
probability with the relevant matter effects.
As we will see, these data will play an important role in the global
fit, since they provide key contributions to the determination of the
atmospheric oscillation parameters and the CP violation phase.
We now discuss them separately.

\subsubsection*{Disappearance channel in MINOS}

The latest measurements of the $\nu_\mu$ disappearance channel in
MINOS have been published in Ref.~\cite{Adamson:2013whj}.  These results
come from the full MINOS data set, collected over a period of nine
years and correspond to exposures of $10.71\times 10^{20}$ protons on
target (POT) in the $\nu_{\mu}$-dominated beam and $3.36\times
10^{20}$ POT in the $\bar{\nu}_{\mu}$-enhanced beam.
One of the key features of these data sample is the preference for a
non-maximal value of the atmospheric mixing angle $\theta_{23}$. In
fact, from the official MINOS analysis, one obtains that maximal
mixing is disfavoured at the 86\% C.L.

\subsubsection*{Appearance channel in MINOS}

The most recent results for the searches of $\nu_e$ appearance in
MINOS have been reported in Ref.~\cite{Adamson:2013ue}. These data
correspond to exposures of $10.6\times10^{20}$ POT in the neutrino
channel and $3.3\times10^{20}$ POT in the anti-neutrino channel.
The neutrino sample is the same as in the preliminary results
presented in the Neutrino-2012 conference, used in our previous
analysis. However, there are some differences in the reconstructed
energy distributions. We are now using the full update from
Ref.~\cite{Adamson:2013ue}.

\subsubsection*{Disappearance channel in T2K } 

The latest results for the $\nu_\mu$ disappearance channel in T2K have
been collected from January 2010 to May 2013, during the four runs of
the experiment and correspond to a total exposure of
$6.57\times10^{20}$ POT~\cite{Abe:2014ugx}.
In comparison with the previous T2K results in
Ref.~\cite{Abe:2013fuq}, sensitivities have been improved thanks to
new event selection and reconstruction techniques, as well as higher
statistics at the near off-axis detector.
A total number of $120$ muon neutrino event candidates have been
observed at the far detector while $446.0\pm 22.5(\text{sys.})$ events
were expected in absence of oscillations.
As we will see in the next section, the T2K disappearance data now
provides the most precise measurement for the atmospheric mixing angle
$\theta_{23}$ with better sensitivity than all other experiments.
Moreover, in contrast to the MINOS $\nu_\mu$ disappearance data, they
prefer a best fit $\theta_{23}$ value very close to maximal. This
point will be crucial for the $\theta_{23}$ octant (in)determination
from the global neutrino oscillation analysis.

\subsubsection*{Appearance channel in T2K } 

As for the disappearance channel, the latest available T2K appearance
data correspond to a total exposure of $6.57\times 10^{20}$ POT,
collected from run 1 to run 4 in the experiment~\cite{Abe:2013hdq}.
A total of $4.92\pm0.55$ background events were expected in the
absence of oscillations, while a sample of 28 electron neutrino events
have been detected. The observed event distribution is consistent with
an appearance signal at $7.3\,\sigma$.

\section{Global fit 2014 results}

In addition to the solar, reactor and long-baseline accelerator
neutrino data described in the previous section, in our global fit to
neutrino oscillations we also include the last results from the
KamLAND reactor experiment presented in Ref.~\cite{Gando:2010aa} as
well as the atmospheric neutrino analysis provided by the
Super-Kamiokande Collaboration in Ref.~\cite{Wendell:2010md}.

\subsection{The role of long--baseline neutrino data in atmospheric
  parameter determination}

\begin{figure}
  \centering
 \includegraphics[width=0.48\textwidth]{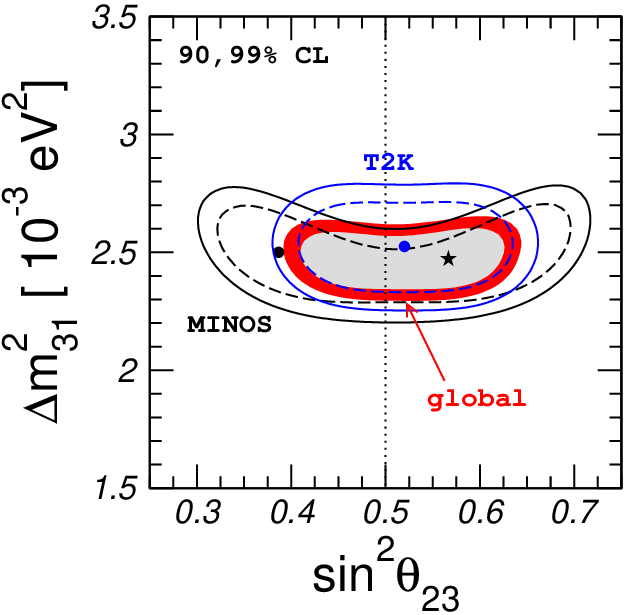}
 \includegraphics[width=0.48\textwidth]{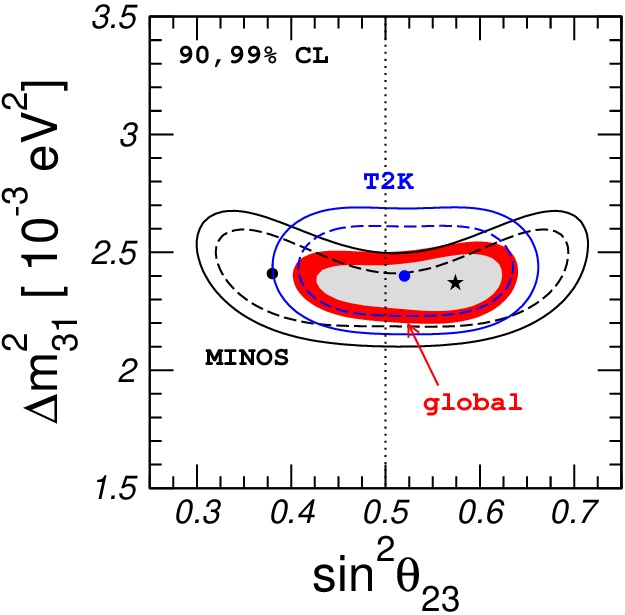}
 \caption{90 and 99\% C.L. regions in the $\sin^2\theta_{23}$ -
   $\Delta m_{31}^2$ plane from separate analysis of MINOS (black
   lines), T2K (blue lines) and from the global analysis of all data
   samples (coloured regions). The left (right) panel corresponds to
   normal (inverted) mass ordering.}
\label{fig:atm-param}
\end{figure}

Long-baseline neutrino data have by now achieved very good
precision. In fact the determination of the atmospheric oscillation
parameters has become fully dominated by the combination of T2K and
MINOS data. This can be appreciated from Fig.~\ref{fig:atm-param},
where one sees how the latest T2K data places the best constraint on
the atmospheric angle $\theta_{23}$, while MINOS still provides the
best determination for the atmospheric mass splitting $\Delta
m^2_{31}$.
Atmospheric neutrino data from Super-Kamiokande are in full agreement
with the parameter regions determined by long-baseline results, though
with less sensitivity.
In this figure we confirm the result obtained by the experimental
collaborations about the maximality of the atmospheric angle. MINOS
data have a mild preference for non maximal $\theta_{23}$, although
$\theta_{23} = \pi/4$ is inside the 90\% CL region for 2 d.o.f. The
absolute best fit point from the analysis of MINOS lies in the first
octant, $\theta_{23} < \pi/4$, although values in the second octant
are allowed with very small $\Delta \chi^2$.
Concerning T2K data, one sees that both for normal (left panel) and
inverted mass hierarchy (right panel) the best fit value is very close
to maximal: $\sin^2\theta_{23} = 0.52$ in both cases, maximal mixing
being allowed with very small $\Delta \chi^2$ with respect to the
absolute minimum: 0.03 (0.02) for normal (inverted) mass ordering.
The global fit preference for values of $\theta_{23}$ in the second
octant emerges after the combination with reactor data, as we will
discuss in the next subsection. We find the best fit points:
\begin{eqnarray}
 \sin^2\theta_{23} & = 0.567^{+0.032}_{-0.128} \quad \Delta m^2_{31} & = (2.48^{+0.05}_{-0.07})\times10^{-3} \rm{eV}^2  \quad (\rm{normal ~ hierarchy}) \\
 \sin^2\theta_{23} & = 0.573^{+0.025}_{-0.043} \quad \Delta m^2_{31} & = (2.38^{+0.05}_{-0.06})\times10^{-3} \rm{eV}^2  \quad (\rm{inverted ~ hierarchy})
 \end{eqnarray}
 Note that for normal hierarchy (NH) a local minimum appears in the first
 octant ($\sin^2\theta_{23} = 0.467$) with $\Delta\chi^2 = 0.28$ with
 respect to the global minimum. For the case of inverted hierarchy (IH),
 solutions with $\sin^2\theta_{23} \leq 0.5$ appear only with
 $\Delta\chi^2 > 1.7$.
 Comparing with our previous global fit, we see that best fit values
 for the atmospheric mixing angle are slightly shifted towards maximal
 values thanks to the latest T2K data. Likewise, $\Delta m^2_{31}$
 values are also shifted towards lower values due to T2K data, which
 now prefer smaller values of the atmospheric mass splitting.

 \subsection{The $\theta_{23}$ octant and the CP violation phase $\delta$}


\begin{figure}
  \centering
 \includegraphics[width=0.75\textwidth]{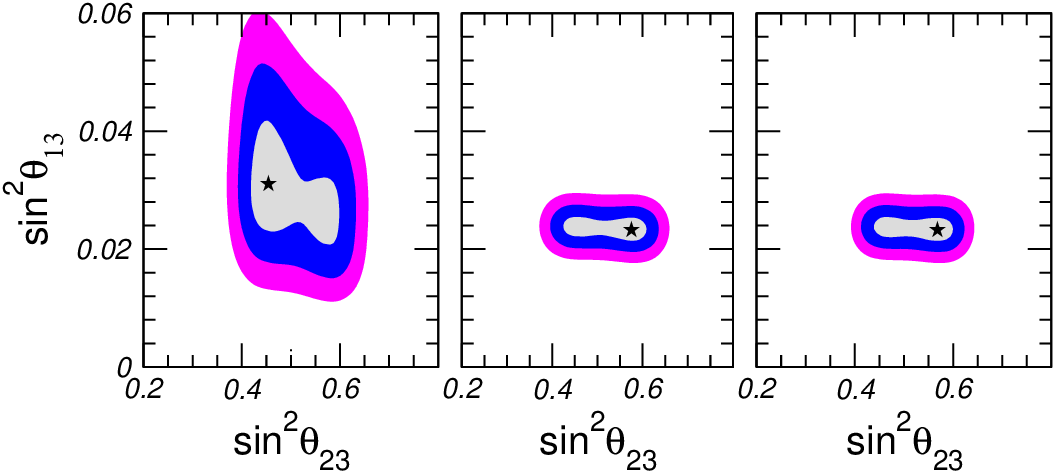}
 \includegraphics[width=0.75\textwidth]{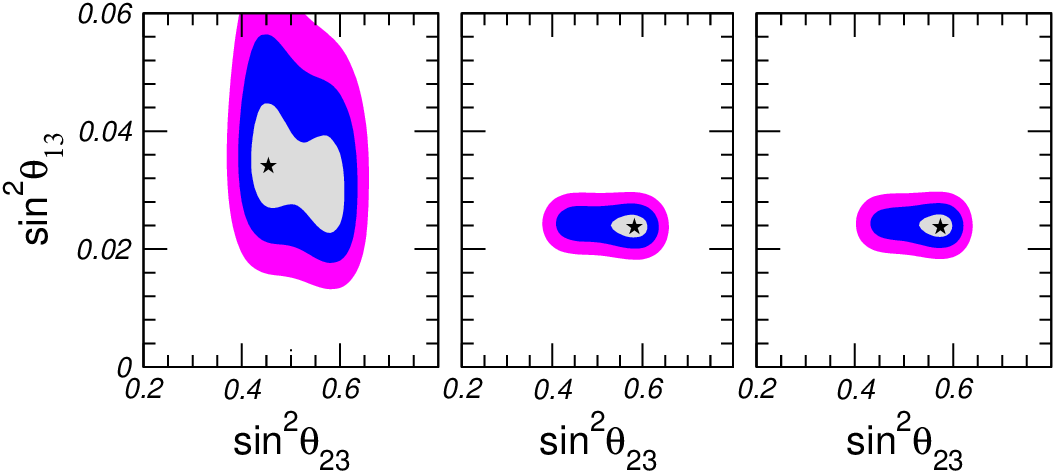}
 \caption{Upper panels: contour regions with $\Delta \chi^2$ = 1, 4, 9
   in the $\sin^2\theta_{23}$ - $\sin^2\theta_{13}$ plane from the
   analysis of long--baseline (MINOS and T2K) + solar + KamLAND data
   (left panel), long-baseline + solar + KamLAND + new Double Chooz,
   Daya Bay and RENO reactor data (middle panel) and the global
   combination (right panel) for normal hierarchy.  Lower panels, the
   same but for inverted neutrino mass hierarchy.}
\label{fig:t23-octant}
\end{figure}

 In this section we will discuss the complementarity between
 long--baseline accelerator and reactor neutrino data in the
 determination of the $\theta_{23}$ octant as well as the CP phase
 $\delta$.
 We will quantify the new sensitivity in the CP violation phase
 $\delta$ as well as the octant of the atmospheric mixing angle
 $\theta_{23}$.
 This emerges by combining the latest accelerator with the latest
 reactor data.

 We start by discussing the effect of the different data samples upon
 the possible preference for a given octant of $\theta_{23}$.  Our
 results are shown in Fig.~\ref{fig:t23-octant}. There we display the
 allowed regions at $\Delta \chi^2$ = 1, 4, 9 in the
 $\sin^2\theta_{23}$ - $\sin^2\theta_{13}$ plane for normal (upper
 panels) and inverted (lower panels) neutrino mass hierarchy. In order
 to appreciate the effect of the individual data sample combinations
 on the parameter determinations we have prepared three different
 panels in this plane. The left panel is obtained by the combination
 of the long-baseline data from MINOS and T2K and the results of all
 solar neutrino experiments plus KamLAND.
 The accelerator MINOS and T2K data already produce a rather
 restricted allowed region in parameter space, showing an
 anti-correlation between $\theta_{23}$ and $\theta_{13}$ coming
 essentially from the oscillation probability in the $\nu_e$
 appearance channel.
In this panel solar and KamLAND impose only minor constraints on the
reactor mixing angle $\theta_{13}$.
In the middle panel of Fig.~\ref{fig:t23-octant}, the data samples
from Double Chooz, Daya Bay and RENO have been included in the
analysis. Here one can see how the very precise determination of
$\theta_{13}$ at reactor experiments, particularly Daya Bay, 
considerably reduces the allowed region. On the other hand, the Daya
Bay preference for values of $\sin^2\theta_{13}$ around 0.023-0.024
moves the best fit value of $\theta_{23}$ to the second octant. This
effect is particularly important for the case of inverted hierarchy,
because of the slightly larger values of $\theta_{13}$ preferred for
$\theta_{23} < \pi/4$. As a result, the first octant region is more
strongly disfavoured so that values of $\sin^2\theta_{23} < \pi/4$ are
allowed only with $\Delta \chi^2 > $ 1.5.
Finally, the right-most panel shows the allowed regions after the
inclusion of the Super-Kamiokande atmospheric
data~\cite{Wendell:2010md}. One can see that there is basically no
change between middle and right panel. This follows from the fact that
the analysis of atmospheric data we adopt does not show a particular
preference for any octant of $\theta_{23}$, both of which are allowed
at 1$\sigma$. This behaviour is also confirmed in the preliminary
versions of updated Super-Kamiokande analysis in
Refs.~\cite{Itow:2013zza,Himmel:2013jva}.

\begin{figure}
  \centering
 \includegraphics[width=0.75\textwidth]{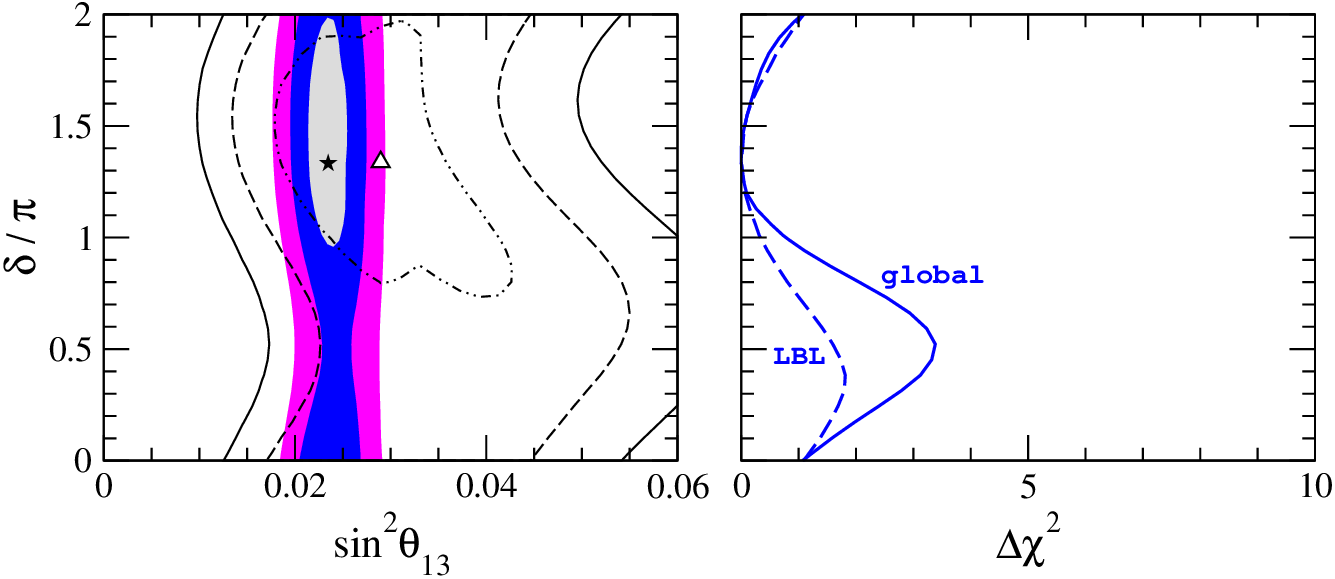}
 \includegraphics[width=0.75\textwidth]{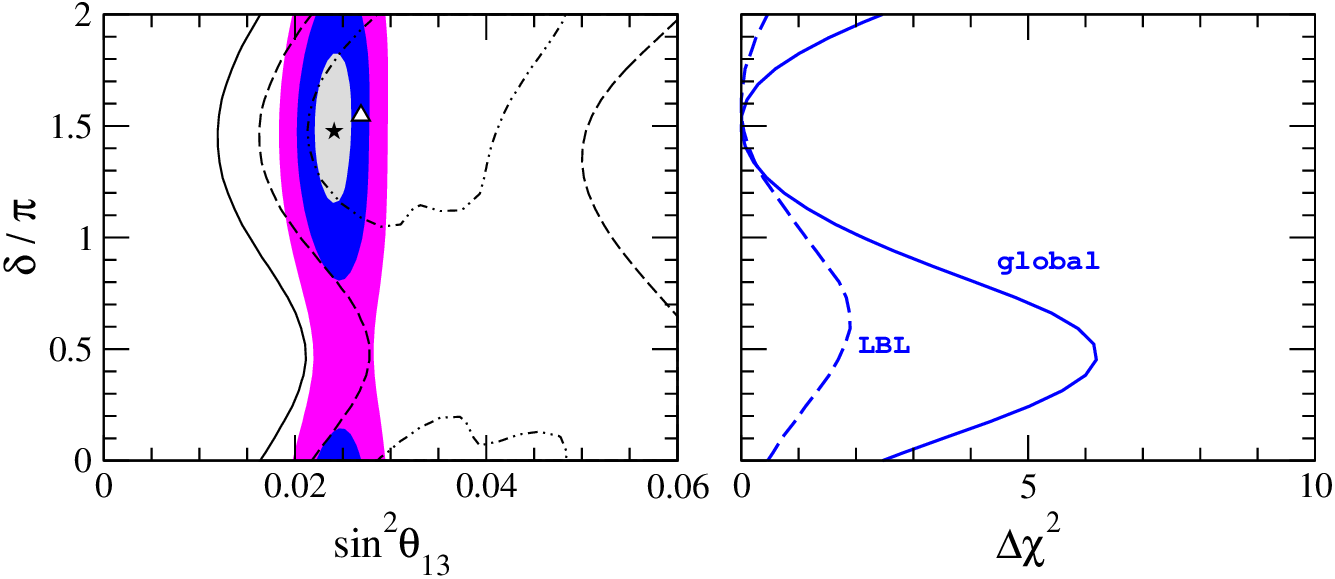}
 \caption{Left panels: contour regions with $\Delta \chi^2$ = 1, 4, 9
   in the $\theta_{13}$-$\delta$ plane from the analysis of LBL data
   alone (lines) and from the combined global analysis (coloured
   regions). Right panels: $\Delta\chi^2$ as a function of the
   CP-violating phase $\delta$ from the analysis of LBL data (dashed
   line) as well as from the global analysis (solid line). Upper
   (lower) figures correspond to NH (IH).}
\label{fig:t13-delta}
\end{figure}

Now we turn to the discussion of the sensitivity to the CP violation
phase, $\delta$.
Our previous global analysis in Ref.~\cite{Tortola:2012te} showed 
essentially no dependence on this phase.
However the new results on $\nu_e$ appearance at long-baseline
experiments in combination with the very precise measurement of
$\theta_{13}$ at reactor experiments provides, for the first time, a
substantial sensitivity to the CP phase $\delta$.
This new effect is illustrated in Fig.~\ref{fig:t13-delta}. Here, left
panels show the allowed regions with $\Delta \chi^2$ = 1, 4, 9 in the
$\sin^2\theta_{13}$-$\delta$ plane from the analysis of long-baseline
accelerator data from MINOS and T2K, in both appearance as well as
disappearance channels. This is indicated by three different line
styles used in the left panels.
On the other hand, the coloured regions correspond to the results
obtained from the global oscillation --analysis.  As expected the
combination with reactor data results in narrower regions for
$\theta_{13}$. One can also notice that there is a mismatch between
the region of $\theta_{13}$ preferred by accelerator data for values
of the CP phase $\delta$ around 0.5$\pi$ and the measured value of
this mixing angle at reactor experiments such as Daya Bay, which
dominates the best fit determination. As a result of this mismatch one
obtains in the global analysis a significant rejection for values of
$\delta$ phase around 0.5$\pi$.
This can be seen in the right panels of Fig.~\ref{fig:t13-delta}. Here
one notices that for normal hierarchy values of $\delta \simeq \pi/2$
are disfavoured with $\Delta\chi^2$ = 3.4 (1.8$\sigma$), while for
inverted hierarchy they are disfavoured with $\Delta\chi^2$ = 6.2
(2.5$\sigma$). In both cases the preferred $\delta$ value is located
close to 1.5$\pi$. The best fit points and 1$\sigma$ errors on
$\delta$ are given by:
\begin{eqnarray}
\delta & = (1.34 ^{+0.64}_{-0.38}) \pi & {\rm (normal ~ hierarchy)}\\
\delta & = (1.48 ^{+0.34}_{-0.32}) \pi & {\rm (inverted ~ hierarchy)}
\end{eqnarray}

 Comparing now with other global neutrino oscillation analyses in the
 literature we find our results on the CP phase qualitatively agree
 with the ones in Refs. \cite{nu-fit, Capozzi:2013csa} for the same
 data included.  Note, however, that the agreement holds for their
 global analysis without atmospheric data, since these authors have
 also included the effect of the $\delta$ in the atmospheric data
 sample, absent in the official Super-Kamiokande analysis adopted
 here. As a result, their global fit results show a somewhat stronger
 rejection against $\delta \simeq \pi/2$ than we find, specially for
 the normal hierarchy case, as expected. In the inverted hierarchy
 case, though, the sensitivity on the CP phase is essentially
 unaffected by the subleading effects on the atmospheric analysis.

\subsection{Summary of global fit}

\begin{figure}
  \centering
 \includegraphics[width=0.8\textwidth]{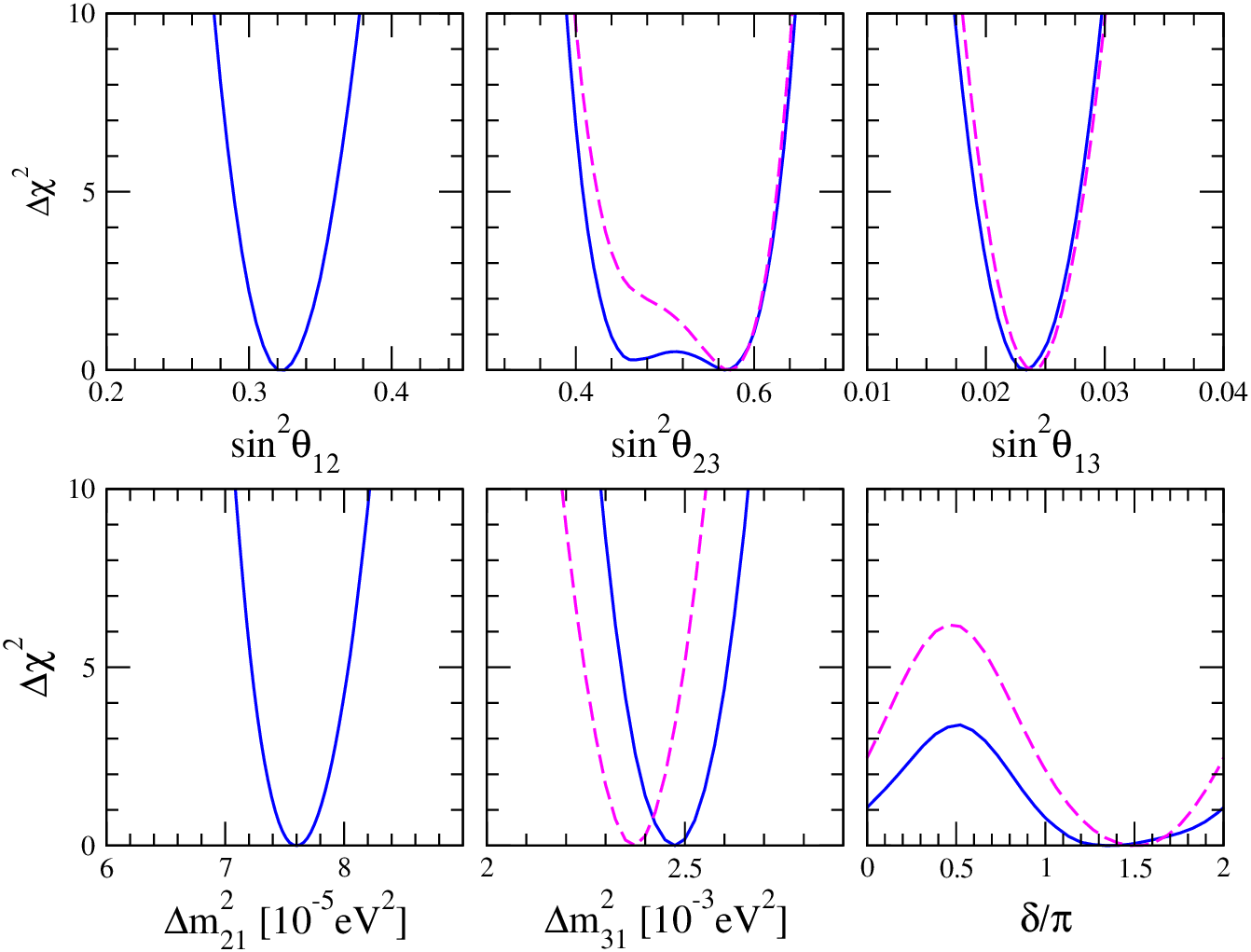}
 \caption{$\Delta\chi^2$ profiles as a function of all the neutrino
   oscillation parameters $\sin^2\theta_{12}$, $\sin^2\theta_{23}$,
   $\sin^2\theta_{13}$, $\Delta m^2_{21}$, $\Delta m^2_{31}$ and the
   CP phase $\delta$. For the central and right panels the solid lines
   correspond to the case of normal mass hierarchy while the dashed
   ones correspond to inverted mass hierarchy.}
\label{fig:summary}
\end{figure}

In this section we summarize the results obtained in our global
analysis to neutrino oscillations.
In Fig.~\ref{fig:summary} we present the $\Delta\chi^2$ profiles as a
function of all neutrino oscillation parameters.  In the panels with
two lines, the solid one corresponds to normal hierarchy while the
dashed one gives the result for inverted mass hierarchy.
Best fit values as well as 1, 2 and 3$\sigma$ allowed ranges for all
the neutrino oscillation parameters are reported in Table
\ref{tab:summary}.

First we note that solar neutrino parameter determination is basically
unchanged with respect to our previous global fit
in~\cite{Tortola:2012te}. We find that the inclusion of the new
SK-IV solar data sample leads only to minor modifications in the
$\sin^2\theta_{12}$ and $\Delta m^2_{21}$ best fit values.
As we already discussed in the previous section, the atmospheric
neutrino parameters are now determined mainly by the new long-baseline
data.
With the new T2K data, the preferred value for the mass splitting
$\Delta m^2_{31}$ is now somewhat smaller, while the best fit value
for the atmospheric angle $\theta_{23}$ has been shifted towards
values closer to maximal. The status of maximal $\theta_{23}$ mixing
angles has also been improved thanks to the latest T2K disappearance
data.
Regarding the reactor mixing angle $\sin^2\theta_{13}$, the more
precise reactor data from Daya Bay and RENO have reduced the allowed
1$\sigma$ range from $\sim$11\% to $\sim$8\%. The preferred value of
$\theta_{13}$ has also been shifted to somewhat smaller values.
Finally, thanks to the combination of the latest accelerator and
reactor neutrino data, we have obtained an enhanced sensitivity to the
CP violation phase. We find preferred values for $\delta$ around $1.5
\pi$ for both mass hierarchies. On the other hand, values close to
$0.5\pi$ are disfavoured at 1.8$\sigma$ (2.5$\sigma$) for normal
(inverted) mass ordering.


\begin{table}[t]\centering
  \catcode`?=\active \def?{\hphantom{0}}
   \begin{tabular}{lccc}
    \hline
    parameter & best fit $\pm$ $1\sigma$ &  2$\sigma$ range& 3$\sigma$ range
    \\
    \hline
    $\Delta m^2_{21}\: [10^{-5}\eVq]$
    & 7.60$^{+0.19}_{-0.18}$  & 7.26--7.99 & 7.11--8.18 \\[3mm] 
    $|\Delta m^2_{31}|\: [10^{-3}\eVq]$ (NH)
    &  2.48$^{+0.05}_{-0.07}$ &  2.35--2.59 &  2.30--2.65\\
    $|\Delta m^2_{31}|\: [10^{-3}\eVq]$ (IH)
    &  2.38$^{+0.05}_{-0.06}$ &  2.26--2.48 &  2.20--2.54 \\[3mm] 
    $\sin^2\theta_{12} / 10^{-1}$
    & 3.23$\pm$0.16 & 2.92--3.57 & 2.78--3.75\\
    $\theta_{12}/\degree$ 
    & 34.6$\pm$1.0 & 32.7--36.7 & 31.8--37.8\\[3mm]  
    $\sin^2\theta_{23} / 10^{-1}$ (NH)
    &	5.67$^{+0.32}_{-1.28}$ \footnote{There  is a local minimum in the first octant, 
        $\sin^2\theta_{23}$ = 0.467 with $\Delta\chi^2 = 0.28$ with respect to the
        global minimum} 
    & 4.13--6.23 & 3.92 -- 6.43 \\
    $\theta_{23}/\degree$ 
    & 48.9$^{+1.9}_{-7.4}$ & 40.0--52.1 & 38.8--53.3 \\ 
    $\sin^2\theta_{23} / 10^{-1}$ (IH)
    & 5.73$^{+0.25}_{-0.43}$ & 4.32--6.21 & 4.03--6.40 \\
    $\theta_{23}/\degree$ 
    & 49.2$^{+1.5}_{-2.5}$ & 41.1--52.0 & 39.4--53.1\\[3mm]  
    $\sin^2\theta_{13} / 10^{-2}$ (NH)
    & 2.34$\pm$0.20 & 1.95--2.74 & 1.77--2.94 \\
    $\theta_{13}/\degree$ 
    &	8.8$\pm$0.4 & 8.0--9.5 & 7.7--9.9\\
    $\sin^2\theta_{13} / 10^{-2}$ (IH)
    & 2.40$\pm$0.19 & 2.02--2.78 & 1.83--2.97 \\
     $\theta_{13}/\degree$ 
     & 8.9$\pm$0.4 & 8.2--9.6 & 7.8--9.9\\[3mm]
   $\delta/\pi$ (NH)
   	& 1.34$^{+0.64}_{-0.38}$ & 0.0--2.0 & 0.0--2.0 \\
   $\delta/\degree$ 
	& 241$^{+115}_{-68}$ & 0--360 & 0--360 \\
   $\delta/\pi$ (IH)	
   	& 1.48$^{+0.34}_{-0.32}$ & 0.0--0.14 \& 0.81-2.0 & 0.0--2.0 \\	
   $\delta/\degree$ 
   	& 266$^{+61}_{-58}$ & 0--25 \& 146--360 & 0--360\\
       \hline
     \end{tabular}
     \caption{ \label{tab:summary} Neutrino oscillation parameters
       summary. For $\Delta m^2_{31}$, $\sin^2\theta_{23}$, $\sin^2\theta_{13}$, 
       and $\delta$ the upper (lower) row corresponds to normal (inverted)
       neutrino mass hierarchy.}
\end{table}

\section{Conclusions}

Here we have updated the global fit of neutrino oscillations given
in~Ref.~\cite{Tortola:2012te} by including the recent measurements of
the last two years. These include the measurements of reactor
anti-neutrino disappearance reported by Daya Bay and RENO, together
with latest long--baseline appearance and disappearance data from T2K
and MINOS. In addition, we have also included the revised data form
the third solar phase of Super-Kamiokande, SK-III, as well as new
solar results from the fourth phase of Super-Kamiokande, SK-IV.
Our results are summarized in the four figures and Table.
We find that for normal mass ordering the global best fit value of the
atmospheric angle $\theta_{23}$ is consistent with maximal mixing at
one-sigma, while for the inverted spectrum maximal mixing appears at
1.3$\sigma$.
We note that the T2K disappearance data now provide the most
sensitive measurement of the atmospheric mixing angle $\theta_{23}$.
We also determine the impact of the new data upon all the other
neutrino oscillation parameters, with emphasis on the increasing
sensitivity to the CP violation phase $\delta$. The latter follows
from the complementarity between accelerator and reactor data and
leads to preferred ranges given in the table.

\section*{Acknowledgments}

This work was supported by the Spanish MINECO under grants
FPA2011-22975 and MULTIDARK CSD2009-00064 (Consolider-Ingenio 2010
Programme).

\section*{APPENDIX: Updated global analysis after Neutrino 2014 conference}

In this Appendix we present an updated global fit after the inclusion
of new data released at the Neutrino 2014 conference in Boston in
June 2014.
We have included the latest data from the reactor experiments Double
Chooz, RENO, and Daya Bay.

Double Chooz presented new results corresponding to 467.9 live
days~\cite{Abe:2014bwa,DChooz:Nu2014}. The analysis of new data, with
twice more statistics than the previous release, improved energy
reconstruction, and lower systematics, implies a slight improvement in
the determination of the reactor mixing angle from the rate + shape
analysis:
\begin{equation}
\sin^2 2\theta_{13}({\rm{Double ~ Chooz}}) = 0.090 ^{+0.032}_{-0.029}
\end{equation}

The RENO experiment released their new rate-only analysis using 800
days of data taking~\cite{RENO:Nu2014}.
\begin{equation}
\sin^2 2\theta_{13}({\rm{RENO}}) = 0.101 \pm 0.008 (\rm{stat}) \pm 0.010 (\rm{syst})
\end{equation}
also improving their former determination of $\theta_{13}$.

Finally, the Daya Bay Collaboration presented their results for a
period of 621 days (four times more statistics than their previous
data release), combining the periods with six and eight antineutrino
detectors~\cite{DayaBay:Nu2014}. From their rate + shape analysis, they
obtain the following best fit value:
\begin{equation}
\sin^2 2\theta_{13}({\rm{Daya ~ Bay}}) = 0.084 \pm 0.005 \, ,
\end{equation}
now determined with an improved 6\% precision, and slightly lower
compared to the previous value.

A distortion in the reactor spectrum in the energy range between 4 and
6 MeV was reported by the RENO and Double Chooz collaborations at the
Neutrino 2014 conference~\cite{DChooz:Nu2014,RENO:Nu2014}. The origin
of this effect, also confirmed by Daya Bay~\cite{DayaBay:ichep}, is
still unknown, although its correlation with the reactor thermal power
indicates it may be consistent with an unaccounted reactor neutrino
flux. In any case, this excess of events around 5 MeV does not affect
the determination of $\theta_{13}$ from reactor experiments, based upon
the comparison of near and far detector rates.

 \subsection{Impact of new reactor data upon the global oscillation fit}

The main difference between this update and the analysis in the previous sections is a slightly 
lower and more precise value of $\theta_{13}$ implied by the recent Daya Bay reactor data:
\begin{eqnarray}
\sin^2 \theta_{13} = &  0.0226 \pm 0.0012  & (\rm {NH})\\
\sin^2 \theta_{13} = &  0.0229 \pm 0.0012  & (\rm {IH}).  
\end{eqnarray}

This result has an impact upon the determination of
$\sin^2\theta_{23}$ through the correlations between these two mixing
angles.
In particular, the slightly lower value of $\sin^2\theta_{13}$ preferred by the 
new Daya Bay data favors values of $\theta_{23}$ in the second octant, 
worsening the status of the first
octant solution. Comparing with the situation before the Neutrino 2014
conference, we find that for NH the local minimum in the first
octant appears now with $\Delta\chi^2$ = 0.36 (vs $\Delta\chi^2$ =
0.28). Concerning the case of IH, we find that values of $\theta_{23}$
in the first octant are allowed only with $\Delta\chi^2 >$ 1.9, to be
compared with $\Delta\chi^2 >$ 1.5 before the inclusion of new reactor
data.

The new determination of $\theta_{13}$ at reactor experiments also has
an impact on the sensitivity to the {\it CP} phase $\delta$ from the global
oscillation analysis.
The lower value implied by new data increases the tension between
long-baseline and reactor data for some values of $\delta$, enhancing
the rejection against $\delta \sim 0.5\pi$. We find that these values are
now disfavored at the 2.0$\sigma$ (2.7$\sigma$) level for normal (inverted)
mass ordering.

\subsection{Summary of the updated global neutrino oscillation analysis}

\begin{figure}
  \centering
 \includegraphics[width=0.8\textwidth]{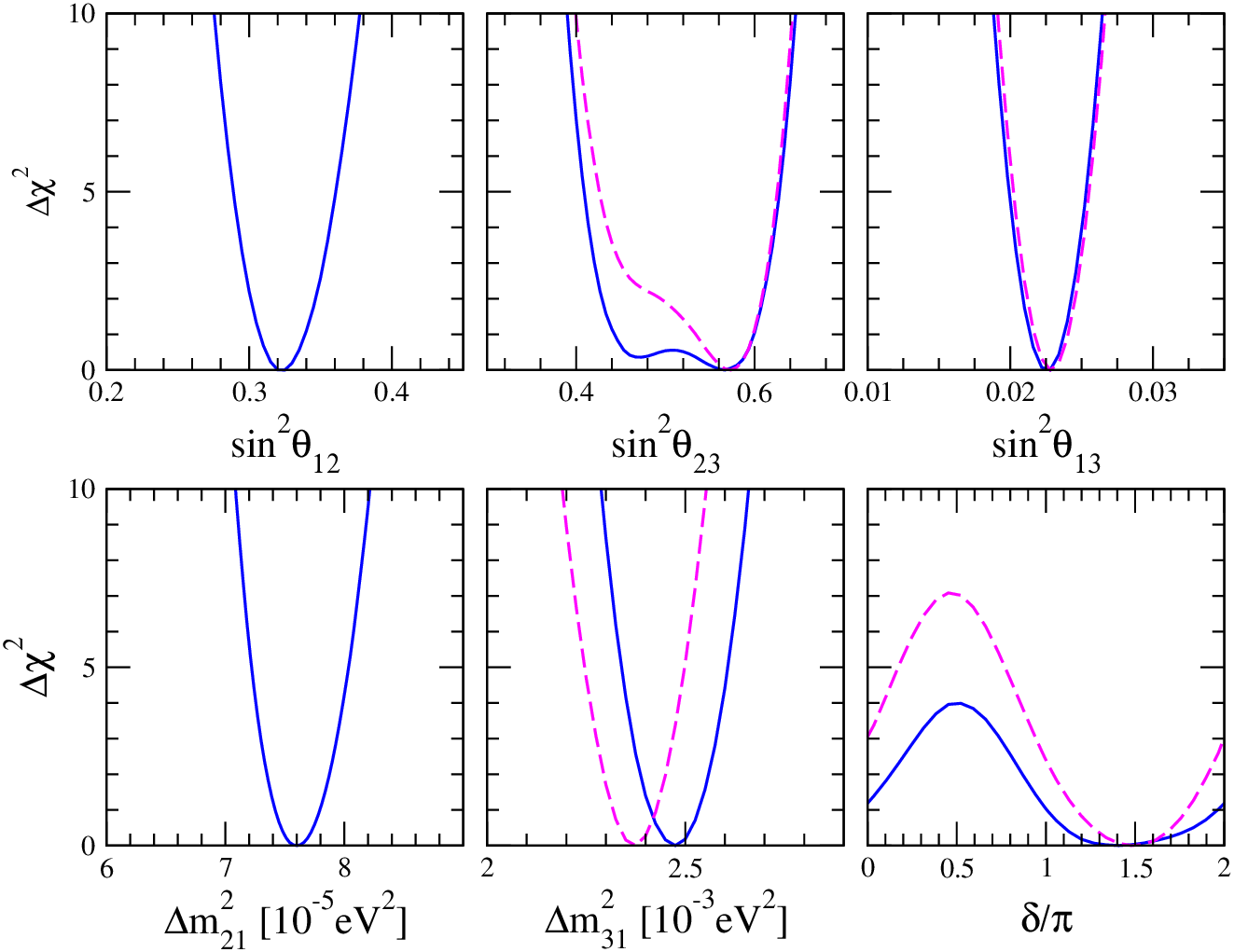}
 \caption{Same as Fig.~\ref{fig:summary} for the updated global
   analysis after Neutrino 2014 conference}
\label{fig:sum-Nu2014}
\end{figure}

The main results of our updated global fit to neutrino oscillations are summarized 
in Fig.~\ref{fig:sum-Nu2014}  and Table \ref{tab:summary-nu14}.
As commented above, the new reactor data presented at the Neutrino 2014 conference 
show a preference for a lower value of $\sin^2\theta_{13}$, now  determined with an accuracy 
of $\sim$ 5\% thanks to the more precise data. 
This modification leaves unchanged all the other oscillation parameters except the atmospheric
mixing angle $\theta_{23}$ and the {\it CP} phase $\delta$.
In both cases, the smaller value of $\theta_{13}$ favored by the new data results in a 
worsening of the solutions already disfavored in our previous analysis, namely, first 
octant solutions for $\theta_{23}$ and values of $\delta \sim 0.5\pi$.

\begin{table}[h!]\centering
  \catcode`?=\active \def?{\hphantom{0}}
   \begin{tabular}{lccc}
    \hline
    parameter & best fit $\pm$ $1\sigma$ &  2$\sigma$ range& 3$\sigma$ range
    \\
    \hline
    $\Delta m^2_{21}\: [10^{-5}\eVq]$
    & 7.60$^{+0.19}_{-0.18}$  & 7.26--7.99 & 7.11--8.18 \\[3mm] 
    $|\Delta m^2_{31}|\: [10^{-3}\eVq]$ (NH)
    &  2.48$^{+0.05}_{-0.07}$ &  2.35--2.59 &  2.30--2.65\\
     $|\Delta m^2_{31}|\: [10^{-3}\eVq]$ (IH)
    &  2.38$^{+0.05}_{-0.06}$  &  2.26--2.48 &  2.20-2.54 \\[3mm] 
    $\sin^2\theta_{12} / 10^{-1}$
    & 3.23$\pm$0.16 & 2.92--3.57 & 2.78--3.75\\
    $\theta_{12} /\degree$
    & 34.6$\pm$1.0 & 32.7--36.7 & 31.8--37.8 \\[3mm]  
    $\sin^2\theta_{23} / 10^{-1}$ (NH)
    &	5.67$^{+0.32}_{-1.24}$ \footnote{There is a local minimum in the first octant, at 
        $\sin^2\theta_{23}$=0.473 with $\Delta\chi^2 = 0.36$ with respect to the
        global minimum} 
    & 4.14--6.23 & 3.93--6.43 \\
    $\theta_{23} /\degree$ 
    &	48.9$^{+1.8}_{-7.2}$ & 40.0--52.1 & 38.8--53.3\\
    $\sin^2\theta_{23} / 10^{-1}$ (IH)
    & 5.73$^{+0.25}_{-0.39}$  & 4.35--6.21 & 4.03--6.40 \\
    $\theta_{23} /\degree$ 
	& 49.2$^{+1.5}_{-2.3}$ & 41.3--52.0 & 39.4--53.1 \\[3mm]  
    $\sin^2\theta_{13} / 10^{-2}$ (NH)
    & 2.26$\pm$0.12 &  2.02--2.50 & 1.90--2.62 \\
    $\theta_{13} /\degree$ 
   & 8.6$^{+0.3}_{-0.2}$ & 8.2--9.1 & 7.9--9.3 \\ 	
    $\sin^2\theta_{13} / 10^{-2}$ (IH)
    & 2.29$\pm$0.12 & 2.05--2.52 & 1.93--2.65 \\
    $\theta_{13} /\degree$ 
    & 8.7$\pm$ 0.2 & 8.2--9.1 &  8.0--9.4 \\[3mm]
   $\delta/\pi$ (NH)
   	& 1.41$^{+0.55}_{-0.40}$ & 0.0--2.0 & 0.0--2.0 \\
   $\delta/\degree$
   	& 254$^{+99}_{-72}$ & 0--360 & 0--360\\ 
    $\delta/\pi$ (IH)	
   	& 1.48$\pm$0.31 & 0.00--0.09 \& 0.86--2.0 & 0.0--2.0 \\
   $\delta/\degree$
	& 266$\pm$56 & 0--16 \& 155--360 & 0--360\\	
       \hline
     \end{tabular}
     \caption{ \label{tab:summary-nu14} Neutrino oscillation
       parameters summary from the global analysis updated after
       Neutrino 2014 conference}
\end{table}

\newpage


\begingroup
\raggedright
\sloppy

\endgroup

\end{document}